\documentclass{iau} 
\usepackage{graphicx}

\providecommand{\e}[1]{\ensuremath{\times 10^{#1}}}
\newcommand{\apj}{\textit{ApJ}}
\newcommand{\apjl}{\textit{ApJL}}
\newcommand{\solphys}{\textit{SoPh}}
\newcommand{\aap}{\textit{A\&A}}
\newcommand{\araa}{\textit{ARA\&A}}
\newcommand{\njop}{\textit{New Journal of Physics}}
\newcommand{\lrsp}{\textit{Living Reviews in Solar Physics}}
\newcommand{\natr}{\textit{Nature}}

\title[Suppression and promotion of flux emergence] %% give here short title %%
{The Suppression and Promotion of Magnetic Flux Emergence in Fully Convective Stars}

\author[Weber et al. 2017]   %% give here short author list %%
{Maria A. Weber, Matthew K. Browning, Suzannah Boardman, Joshua Clarke, Samuel Pugsley, Edward Townsend}
\affiliation{Department of Physics and Astronomy, University of Exeter, \\ Stocker Road, EX4 4QL Exeter, UK \\ email: {\tt mweber@astro.ex.ac.uk}}

\pubyear{2016}
\volume{328}  %% insert here IAU Symposium No.
\jname{Living Around Active Stars}
\editors{D. Nandi, ed.}
\begin{document}

\maketitle

\begin{abstract}
Evidence of surface magnetism is now observed on an increasing number of cool stars. The detailed manner by which dynamo-generated magnetic fields giving rise to starspots traverse the convection zone still remains unclear. Some insight into this flux emergence mechanism has been gained by assuming bundles of magnetic field can be represented by idealized thin flux tubes (TFTs). \cite{wb2016} have recently investigated how individual flux tubes might evolve in a 0.3M$_{\odot}$ M dwarf by effectively embedding TFTs in time-dependent flows representative of a fully convective star. We expand upon this work by initiating flux tubes at various depths in the upper $\sim$50-75$\%$ of the star in order to sample the differing convective flow pattern and differential rotation across this region. Specifically, we comment on the role of differential rotation and time-varying flows in both the suppression and promotion of the magnetic flux emergence process.

\end{abstract}

\firstsection % if your document starts with a section,
\section{Introduction}
M dwarfs are among the most magnetically active and abundant stars in the galaxy. They encompass a broad mass range of $\sim$0.08-0.6M$_{\odot}$. It is widely thought that the seat of the dynamo in partially convective dwarfs ($\ge$0.35M$_{\odot}$) resides in the tachocline, a region of shear at the interface between the convection and radiative zones (e.g. \cite[Charbonneau 2010]{charlrsp2010}). In this interface dynamo, the toroidal magnetic field is amplified in the tachocline and then rises to the surface where it may be observed as starspots and serve as the launching site for strong flares. Yet M dwarfs on the fully convective side of the `tachocline divide' ($\leq$0.35M$_{\odot}$) still effectively build magnetic fields, with this activity increasing in prevalence toward late M spectral types (e.g. \cite[West et al. 2015]{west2015}). This magnetism is similar to that observed in solar-like stars; i.e., there is still a rotation-activity correlation that plateaus at rapid rotation (e.g. \cite[Wright \& Drake 2016]{wright2016}).    

Many have turned to global-scale dynamo models as a way of self-consistently capturing the strength and morphology of magnetism that may be built in fully convective stars (e.g. \cite[Dobler et al. 2006]{dobler2006}; \cite[Browning 2008]{browning2008}; \cite[Yadav et al. 2015]{yadav2015}). Similar simulations are just now beginning to capture some aspects of magnetic flux emergence in rapidly rotating Suns, exhibiting buoyant magnetic loops that emerge naturally from wreaths of magnetism (e.g. \cite[Nelson et al. 2011]{nelson}). But, these simulations are remarkably expensive to compute. As an alternative and less computationally expensive method, the flux tube model has been invoked to describe the evolution of magnetic field bundles, utilizing either the effectively 1D thin flux tube (TFT) approximation or solving the full 3D magnetohydrodynamic (MHD) equations (see e.g. review by \cite[Fan 2009]{fan2009}). Applying this model to the Sun has provided a wealth of insight regarding the flux emergence process, replicating many observed features of active regions (e.g. \cite[Fan 2009]{fan2009}; \cite[Weber et al. 2013]{weber2013}). These simulations assume the dynamo has already built fibril magnetic flux tubes that rise under the combined effects of buoyancy and advection by turbulent flows.    

\cite{wb2016} (hereafter WB16) recently investigated for the first time how flux tubes in a fully convective M dwarf might rise under the joint effects of buoyancy, differential rotation, and convection. The work presented here expands upon the parameter space explored in WB16 by initiating flux tubes at multiple depths between 0.475-0.75R to sample the varying convective flow field structure across this region. In Section \ref{sec:formulation}, we introduce our model and initial conditions. We present the results of some new TFT simulations and diagnostic metrics in Section \ref{sec:results}. We focus on how convection modulates the initially toroidal flux tube (Sec. \ref{sec:conv_effects}), the way differential rotation and time-varying flows may alter the duration and trajectory of the flux tube rise (Sec. \ref{sec:rise_and_lats}), and discuss how convection can suppress the mean rise of magnetism (Sec. \ref{sec:mag_pump}).            

\section{Formulating the Problem}
\label{sec:formulation}
The dynamic evolution of isolated, fibril magnetic flux tubes can be modeled by applying the thin flux tube (TFT) approximation (e.g., \cite[Roberts \& Webb 1978]{roberts1978}; \cite[Spruit 1981]{spruit1981b}). Derived from ideal MHD, the TFT approximation assumes the diameter of the flux tube is small enough that all variables can be represented by their averages over a tube segment, reducing the model to one dimension. To capture the effects of global-scale convection on flux emergence in a 0.3M$_{\odot}$ fully convective star, a convective velocity field computed apart from the TFT models is incorporated through the aerodynamic drag force influencing each flux tube segment. A description of the equations and methods used to solve for the evolution of individual flux tubes is discussed in \cite[WB16]{wb2016}.  

The traditional TFT approach assumes that the magnetic field is generated in the tachocline region of a solar-like star through an interface dynamo.  Here we postulate that a distributed dynamo might build toroidal flux tubes as well. We adopt the simplifying assumption that turbulent, large-scale convective motions have built such flux tubes initially co-rotating with the local differential rotation and in thermal equilibrium with the background fluid. The condition of thermal equilibrium renders the flux tube initially buoyant with a density deficit of $\rho_{e}-\rho = (\rho_{e}B_{0}^{2})/(8 \pi p_{e})$. 

Flux tubes are initiated at latitudes of $1^{\circ}$-$60^{\circ}$ in both hemispheres and depths between 0.475-0.75R in order to sample the varying differential rotation and convective pattern with depth. In this investigation, we are interested in studying the evolution of magnetic flux tubes that might produce starspots. To parallel our work in \cite[WB16]{wb2016}, we perform simulations where $B_{0}$=30-200 kG and the cross-sectional radius $a$ of each flux tube is 1.7\e{8} cm. Assuming the constant total flux of the tube is given by $\Phi=B \pi a^{2}$, the magnetic flux spans from 2.72\e{21} Mx for 30 kG tubes to 1.82\e{22} Mx for 200 kG tubes. This range of $\Phi$ is typical of solar active regions (e.g. \cite[Zwaan 1987]{zwaan_1987}). As the flux tube approaches the surface, the cross-section expands quickly due to the rapid decrease of the density and pressure of the external plasma, and the TFT approximation is no longer satisfied. Therefore, we terminate our simulations once the fastest rising portion of the tube reaches $0.95R$, assuming the motion through the remaining $0.05R$ is negligible.   

\begin{figure}
\begin{center}
 \vspace{0.0\textwidth}   % Shift close to the panel top 
     \centerline{\normalsize     % Includes the labels (here needs the color 
                                %   package, see beginning of this file)
      \hspace{0.0 \textwidth}  {(a)}
      \hspace{0.17\textwidth}  {(b)}
      \hspace{0.42 \textwidth} {(c)}
         \hfill}
     \vspace{-0.04\textwidth}    % Shift back to the panel bottom 
     
 \includegraphics[scale=.35,clip=true,trim=.3cm 1.3cm .6cm 0cm]{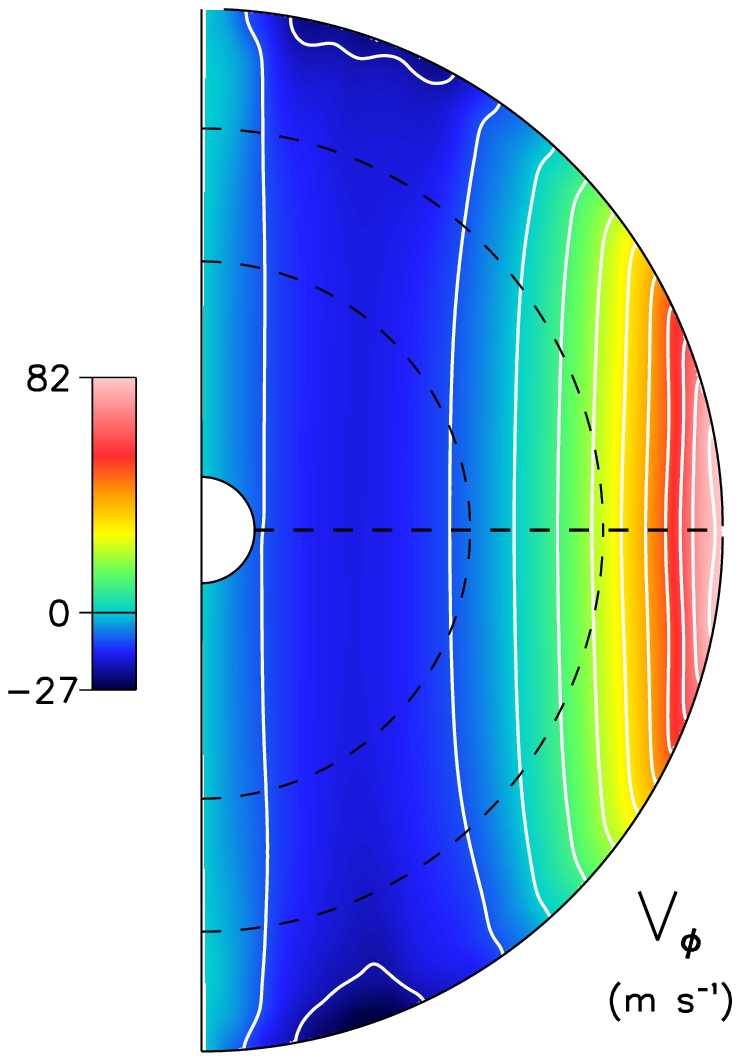} 
 \includegraphics[scale=.3,clip=true,trim=.3cm 0cm .3cm .8cm]{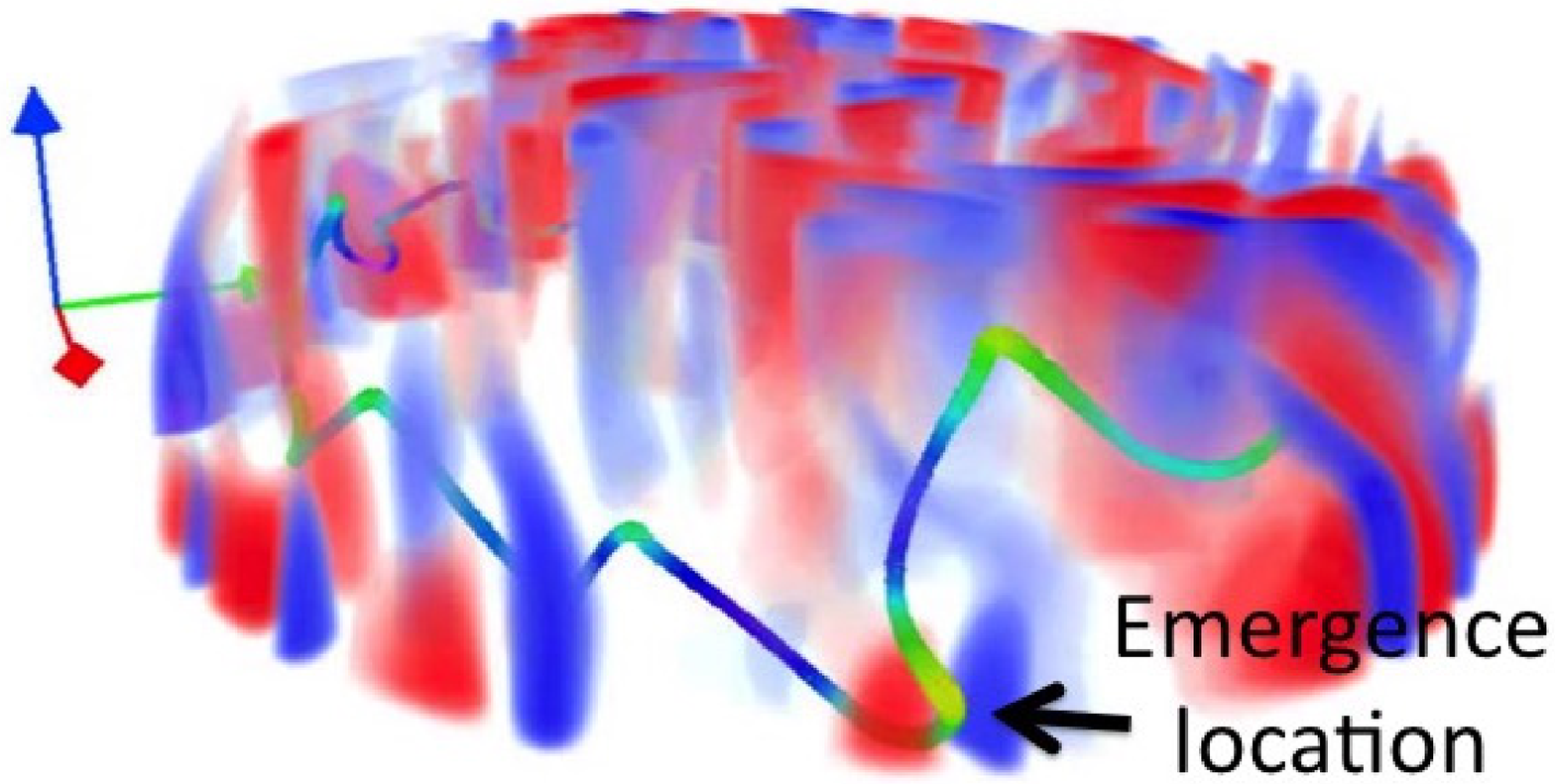}
 \includegraphics[scale=.72,clip=true,trim=2.2cm 3.5cm 2.2cm 2cm]{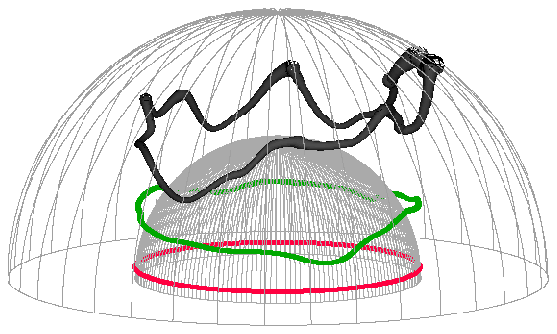}

 \caption{Convective velocity field and representative flux tubes advected by fluid flows in a fully convective star. (a) Meridional plot of the longitudinal velocity $\hat{v}_{\phi}$ relative to the rotating frame averaged over $\sim$460 days, with contour intervals every 10 m s$^{-1}$. Dashed lines at 0.5 and 0.75R. (b) Snapshot of a 30 kG flux tube initiated at $5^{\circ}$ and 0.75R. Tube emerges in a region of strong upflow (red in online version) next to a strong downflow lane (blue), and is colored according to its magnetic field strength, with darker/lighter tones representing stronger/weaker fields. Only the northern hemisphere from the equator to half the stellar radius in the vertical direction is shown. The tube is rendered with a 3D extent according to the local cross-sectional radius. (c) Snapshots in time of a 30 kG flux tube initiated at $5^{\circ}$ and 0.5R. Inner/outer mesh spheres represent surfaces of constant radius at $r_{0}$/0.95R. Shown is the initial flux tube (red), 75$\%$ in time through the total rise (green), and once reaching 0.95R (black). Evolution of this tube is largely parallel to the rotation axis, unlike the more radial rise of the tube apex in (b).}
\label{fig:flow_and_tube}
\end{center}
\end{figure}

To model the external velocity field, we utilize the Anelastic Spherical Harmonic (ASH) code, which solves the 3D MHD equations under the anelastic approximation. Representative of fluid motions in a low-mass, fully convective star, this hydrodynamic ASH simulation captures differential rotation and giant-cell convection in a rotating spherical domain ($\Omega_{0}$=2.6\e{-6} rad s$^{-1}$) spanning from 0.10 to 0.97R, with R the stellar radius of 2.013\e{10} cm (similar to Case C in \cite[Browning 2008]{browning2008}). In \cite[WB16]{wb2016}, we examine how the evolution of flux tubes may change when subject to varying differential rotation profiles. Here we use the differential rotation profile with an angular velocity contrast at the surface between the equator and $60^{\circ}$ of $\Delta \Omega/\Omega_{0}$$\sim$$22\%$ (Fig. \ref{fig:flow_and_tube}a). An instantaneous view of the radial velocity field is shown in Figure \ref{fig:flow_and_tube}b.

To sample the time-varying velocity field in a uniform way, we perform ensemble simulations.  Flux tubes in each ensemble are initialized at the same moment, but evolve independently of each other. Therefore, they are advected by the exact same time-varying convective flows. The simulations performed here align with Cases TLf /TLfC in \cite[WB16]{wb2016}, but only comprise of one ensemble as compared to the three performed in that paper.   

\section{Results}
\label{sec:results}
\subsection{The dynamic evolution of flux tubes and modulation by convective motions}
\label{sec:conv_effects}
In the quiescent interior of a 0.3M$_{\odot}$ star, flux tubes in an axisymmetric configuration in thermal equilibrium with the background plasma rise with a trajectory largely parallel to the rotation axis. Explored in detail in \cite[WB16]{wb2016}, this motion is governed by the force balance the tube achieves in the direction perpendicular to the rotation axis. The four main forces that contribute to flux tube evolution include: buoyancy, magnetic tension, the Coriolis force, and aerodynamic drag. In the early stages, the flux tube adjusts until there is an equilibrium of the inward and outward-directed forces (toward/away from rotation axis). Once reaching this equilibrium, the tube rises parallel to the rotation axis due in large part to the unbalanced vertical component of the buoyancy force. 

%Flux tubes of larger magnetic field strength will move further outward from the rotation axis before reaching equilibrium owing to a larger radially-directed buoyancy force compared to the inward-directed magnetic tension. For the same $B_{0}$, tubes in deeper regions will experience a relative increase of the magnetic tension compared to the buoyancy force.

Radial convective motions distort the shape of the toroidal ring, promoting buoyantly rising loops (Figs. \ref{fig:flow_and_tube}b-c). Strong downflows pin parts of the flux tube in deeper layers, while strong upflows may boost loops toward the surface. By the time the fastest-rising peak (i.e. flux tube apex) reaches the upper boundary, the tube has developed a number of loops with troughs that stretch a large fraction of the star. For flux tubes of similar thickness (i.e. similar $a$), the modulation by convective flows increases with decreasing magnetic field strength. The 30 kG tubes in Figures \ref{fig:flow_and_tube}b-c are strongly advected by convection. In comparison to tubes of stronger magnetic field strength, those of weaker $B_{0}$ have relatively smaller magnetic tension and buoyancy forces, rendering them more susceptible to the aerodynamic drag imparted by turbulent flows. Conversely, tubes of larger $B_{0}$ evolve more like the flux rings in a quiescent interior described above.    

\subsection{Rise times, emergence latitudes, and the influence of convective flows}
\label{sec:rise_and_lats}
\begin{figure}
\begin{center}

 \vspace{0.0\textwidth}   % Shift close to the panel top 
     \centerline{\normalsize     % Includes the labels (here needs the color 
                                %   package, see beginning of this file)
      \hspace{0.04 \textwidth}  {(a)}
      \hspace{0.265\textwidth}  {(b)}
      \hspace{0.27 \textwidth} {(c)}
         \hfill}
     \vspace{-0.03\textwidth}    % Shift back to the panel bottom 

 \includegraphics[scale=.45,clip=true,trim=.03cm 1cm .05cm 0cm]{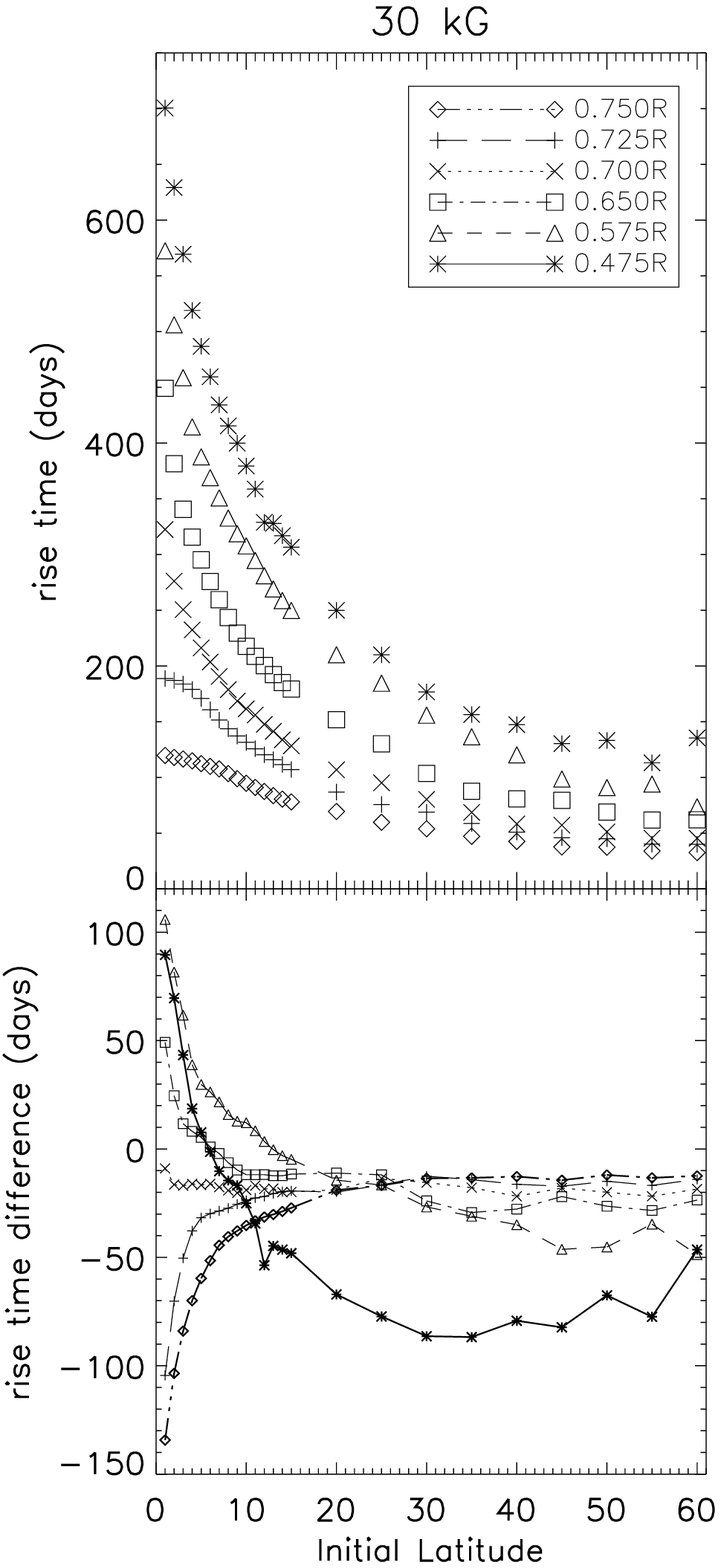} 
 \includegraphics[scale=.45,clip=true,trim=.8cm 1cm .05cm 0cm]{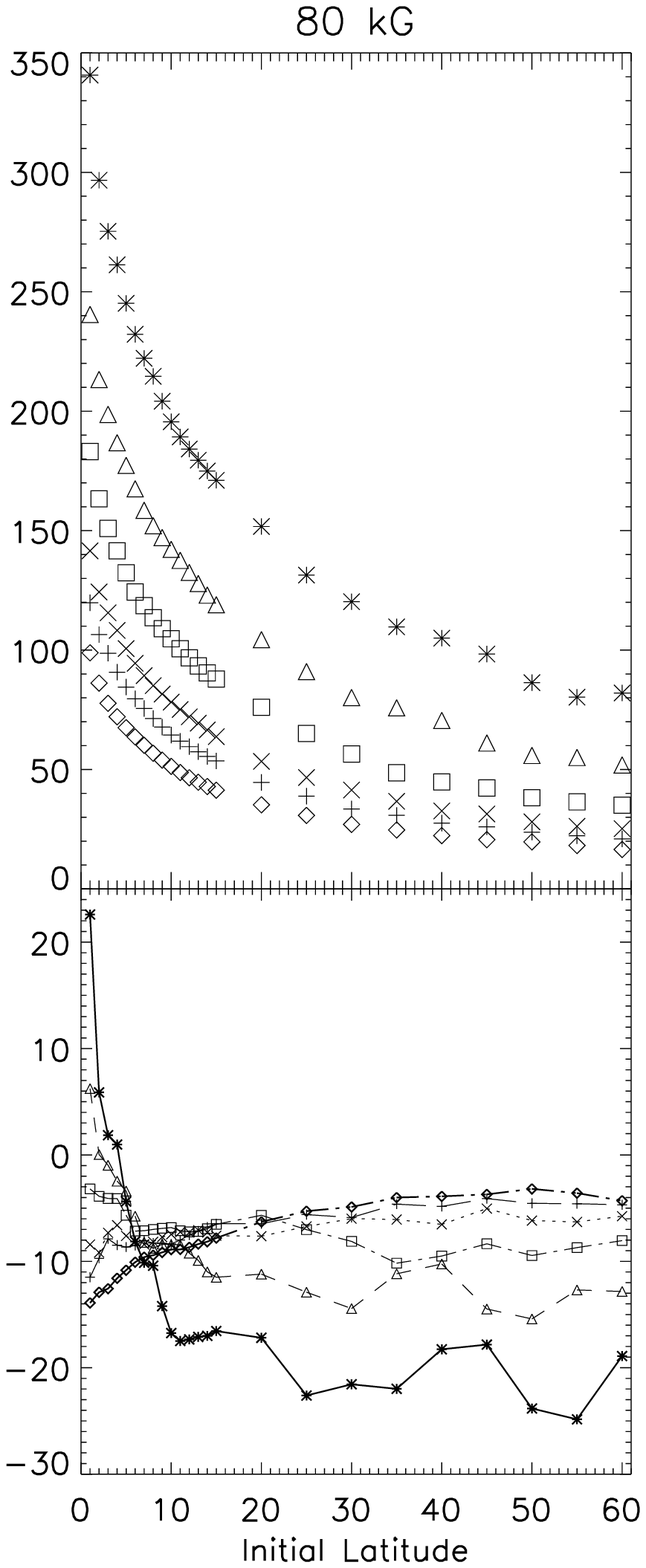}
 \includegraphics[scale=.45,clip=true,trim=.8cm 1cm .05cm 0cm]{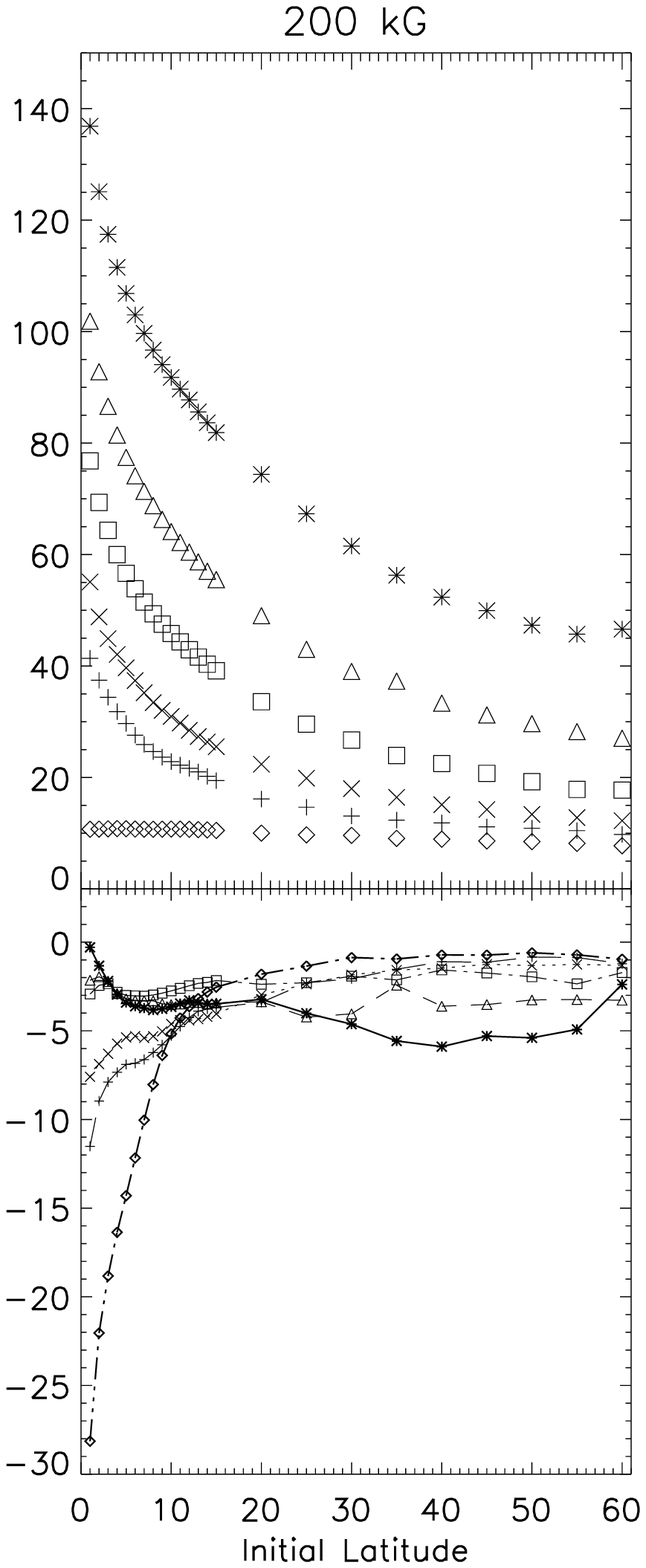}

 \caption{Average apex rise duration ($t_{apex}$) for flux tubes rising subject to flow fields in a fully convective star (top panels) as a function of $|\theta_{0}|$. The difference in rise times for tubes with the same initial conditions rising in a quiescent interior ($t_{q}$) and those rising through the convective flow field is shown below ($t_{apex}-t_{q}$). A positive value for this quantity indicates that the tube apex rises slower than the same tube in a quiescent interior. This scenario is more likely to occur for weaker $B_{0}$ tubes originating at lower latitudes in deeper layers. Quantities are plotted for tubes originating at depths ranging from 0.475-0.750R. Each plot has a different y-axis range.}
   \label{fig:rise_times}
\end{center}
\end{figure}

Mean and local flows can significantly alter the duration and trajectory of a buoyantly rising loop's journey to the stellar surface. Figure \ref{fig:rise_times} shows the total rise time of the flux tube apex for those that evolve subject to convection. Also shown is the relative difference in rise times between tubes with the same initial conditions that rise both with and without convective effects. Each symbol in all the plots in this paper represent the average for flux tubes originating with the same $|\theta_{0}|$. Within the parameter space explored, rise times vary greatly from more than 600 days (30 kG, deeper interior) to as fast as 10 days (200 kG, nearer surface). Generally, flux tubes of a stronger magnetic field strength rise more rapidly because of their greater buoyancy. Tubes initiated at higher latitudes also tend to rise more quickly. This is driven in part by a larger poleward acceleration arising from the magnetic tension force due to a smaller radius of curvature there. Also, from a geometric standpoint, tubes initiated at higher latitudes have a trajectory parallel to the rotation axis that may cover a shorter distance across the convection zone.

A few trends emerge in the bottom panels of Figure \ref{fig:rise_times}. One is the tendency for the duration of the apex rise to roughly follow the rise of the flux rings that evolve without convection. Yet, weaker 30 kG flux tubes initiated at lower latitudes between 0.475-0.65R rise slower than the same flux tubes rising through a quiescent convection zone. This is indicative of a process akin to magnetic pumping (see Sec. \ref{sec:mag_pump}). Another striking feature is the faster relative rise time of especially the 30 kG and 80 kG flux tubes initiated at latitudes $\gtrsim10^{\circ}$ in the deep interior at 0.475R. It is not the case that the whole tube rises faster than the identical tube in a quiescent interior. Rather, the fastest rising loop of the flux tube evolves in such a way that its journey to the surface is boosted from interaction with the mean and time-varying flows. At such depths, the differential rotation is more strongly retrograde (see Fig. \ref{fig:flow_and_tube}a), and may be the reason for the faster apex rise time rather than a change in the nature of the giant-cell structure.  

\begin{figure}
\begin{center}

\vspace{0.0\textwidth}   % Shift close to the panel top 
     \centerline{\normalsize     % Includes the labels (here needs the color 
                                %   package, see beginning of this file)
      \hspace{0.035 \textwidth}  {(a)}
      \hspace{0.435\textwidth}  {(b)}
         \hfill}
     \vspace{-0.03\textwidth}    % Shift back to the panel bottom

 \includegraphics[scale=.513,clip=true,trim=1cm .5cm .7cm 0cm]{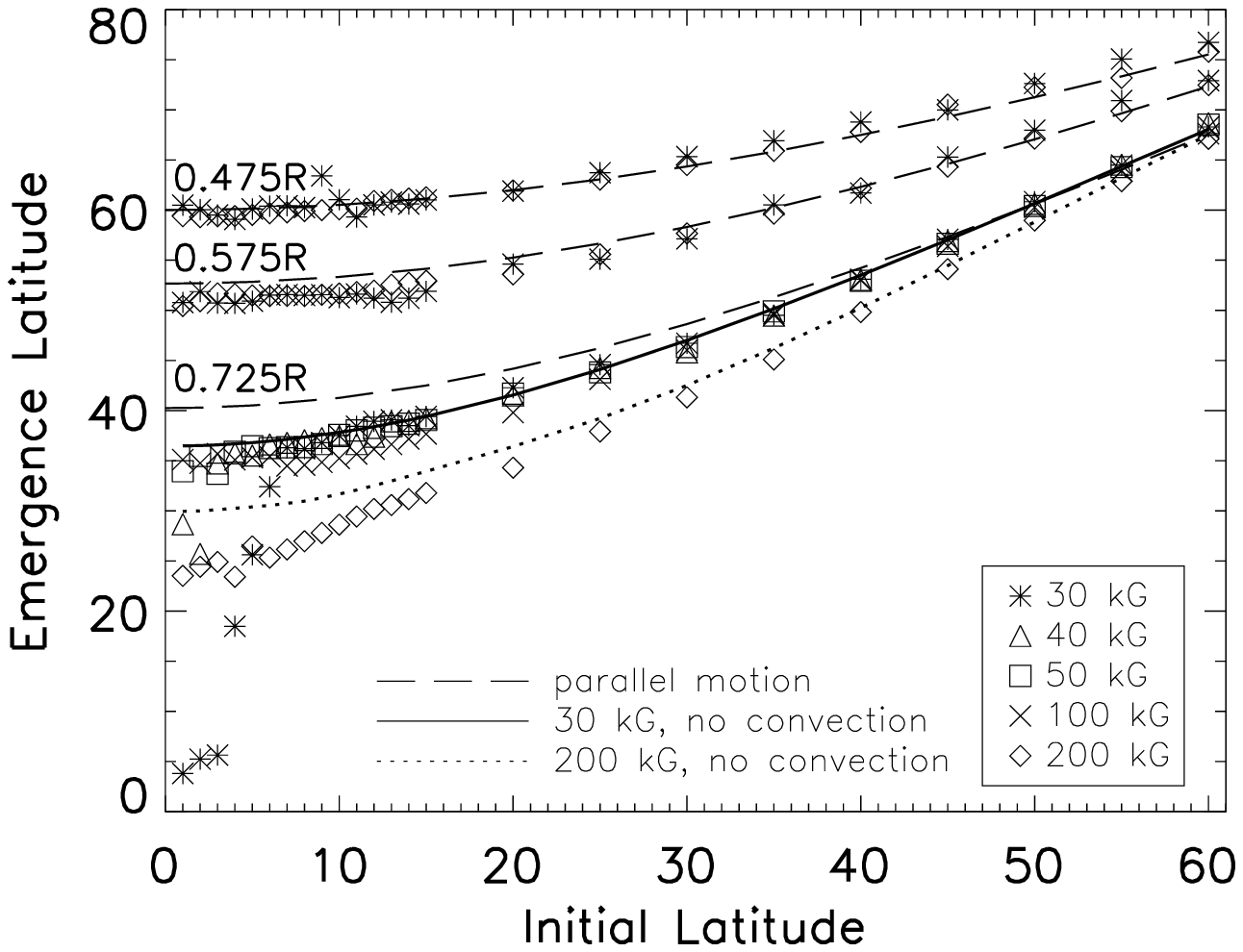} 
 \includegraphics[scale=.513,clip=true,trim=2cm .5cm .7cm 0cm]{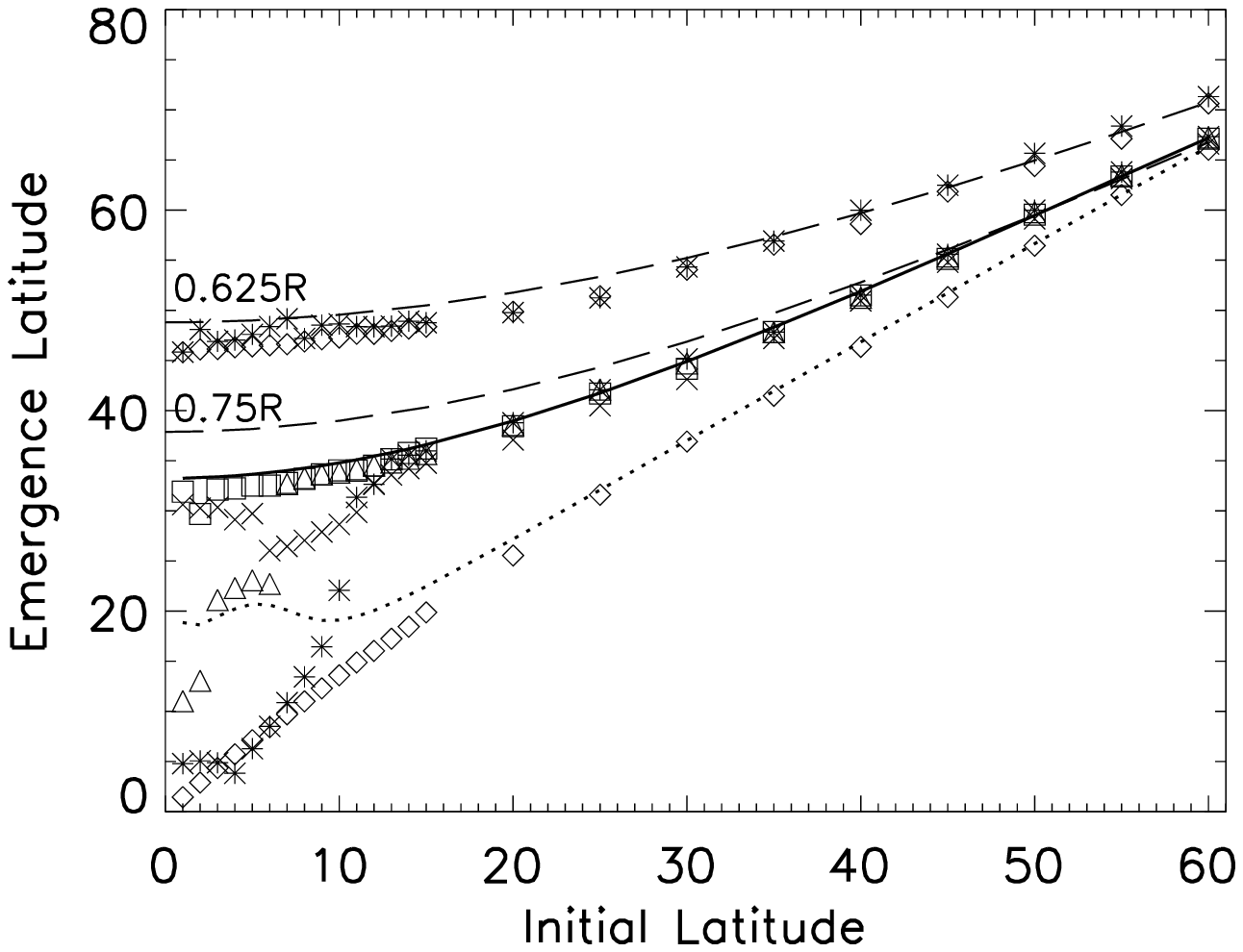}

 \caption{Average emergence latitude of the flux tube apex as a function of $|\theta_{0}|$. These are shown for flux tubes of various $B_{0}$ originating at (a) 0.475, 0.575, and 0.725R and (b) 0.625 and 0.750R. Dashed lines represent the emergence latitude expected if the tube were to rise exactly parallel to the rotation axis. Curves depicted in the legend correspond to tubes rising through a quiescent interior, only shown for those originating at 0.725 and 0.750R. Large deviations from parallel motion for 30-40 kG tubes initiated in shallower regions nearer the equator is largely a result of the strong prograde differential rotation there, suppling angular momentum to these weaker $B_{0}$ loops that are strongly advected by convection. The deviation for the 200 kG tubes in shallower regions is related to the initial force imbalance perpendicular to the rotation axis and vigorous time-varying flows closer to the surface.}
   \label{fig:em_lats}
\end{center}
\end{figure}
%Some of the tubes represented here cover a different parameter space than Figs. \ref{fig:rise_times}, \ref{fig:rbavg}, and \ref{fig:deltar}.

Initially low latitude flux tubes of 30 kG and 200 kG originating between 0.725-0.75R show a significant decrease in their rise times relative to the corresponding flux tubes that rise without convection. These same tubes also have apices that rise more radially, in stark contrast to the parallel trajectories exhibited by the majority of flux tubes we study here (see Fig. \ref{fig:em_lats}). Explained in detail in \cite[WB16]{wb2016}, both of these effects are facilitated by the strongly prograde differential rotation in the upper $\sim$30$\%$ of the convection zone at lower latitudes. If allowed to continue evolving after the apex reaches 0.95R, the same flux tube might produce multiple starspot regions. For example, the flux tube in Figure \ref{fig:flow_and_tube}c initiated at 0.5R could go on to produce three or more high latitude starspots. The tube in Figure \ref{fig:flow_and_tube}b initiated at 0.75R could give rise to at least one near-equatorial spot, but may also potentially produce a few higher latitude spots.

\subsection{Suppression of flux emergence by convection}
\label{sec:mag_pump}

The flux tube properties discussed in Section \ref{sec:rise_and_lats} are representative of the apex of the fastest rising loop.  However, a large fraction of the flux tube may in fact remain in deeper layers (see Figs. \ref{fig:flow_and_tube}b-c).  Through interaction with a sequence of favorable flows, a buoyant loop may reach the surface faster than the commensurate flux tube rising through a quiescent interior. On the other hand, the journey of a rising loop may be prolonged compared to the quiescent case due to buffeting of the flux tube by downflows.  

As in \cite[WB16]{wb2016}, to assess the ability of convection to suppress the mean rise of the flux tube, we compute the average magnetic field weighted radial depth of the flux tube     
\begin{equation}
\langle r(t) \rangle =\frac{\int_{u_{0}}^{u_{N-1}} r(u,t) B(u,t) du }{\int_{u_{0}}^{u_{N-1}} B(u,t) du },
\end{equation} 
where $u_{j}=s/L=j/(N-1)$ for $j=0,..,N-1$ is the fractional arc length along the flux tube, $s$ is the length of the tube up to a mesh point $j$ from the origin, $L$ is the total flux tube length, and $N$ is the number of uniformly spaced mesh points.  The quantity $\langle r(t) \rangle$ represents the radial depth where the majority of the flux tube magnetic field resides.  It places a reduced weight on the magnetic field in shallower layers, which has diminished in strength as portions of the tube rise and subsequently expand.  Similar treatments have been employed for magnetic fields in 3D computational domains (e.g. \cite[Tobias et al. 2001]{tobias2001}; \cite[Abbett et al. 2004]{abbett2004}).

\begin{figure}
\begin{center}
 
 \vspace{0.0\textwidth}   % Shift close to the panel top 
     \centerline{\normalsize     % Includes the labels (here needs the color 
                                %   package, see beginning of this file)
      \hspace{0.035 \textwidth}  {(a)}
      \hspace{0.435\textwidth}  {(b)}
         \hfill}
     \vspace{-0.03\textwidth}    % Shift back to the panel bottom
 
 \includegraphics[scale=.509,clip=true,trim=1cm .5cm .7cm 0cm]{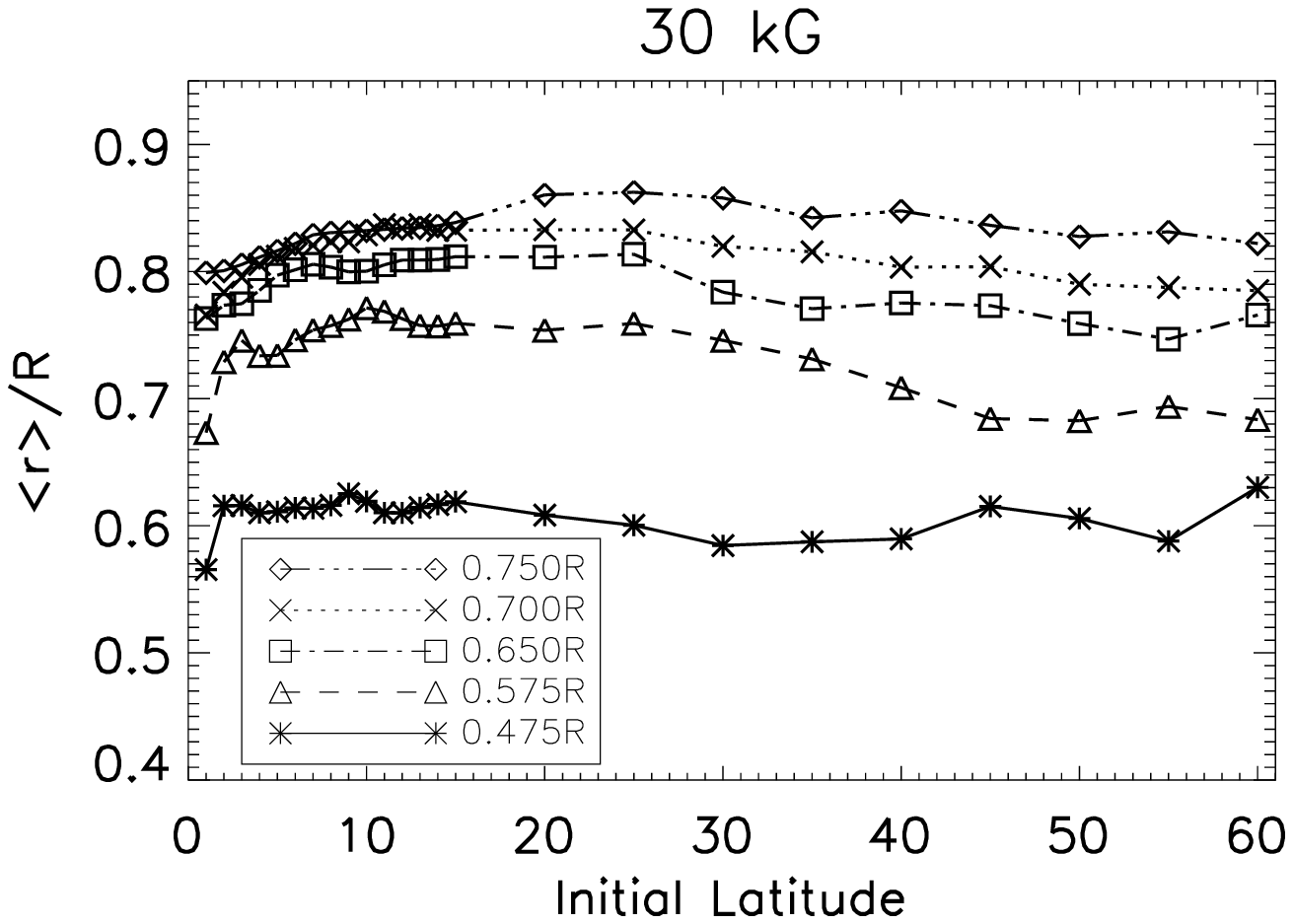} 
 \includegraphics[scale=.509,clip=true,trim=1.8cm .5cm .7cm 0cm]{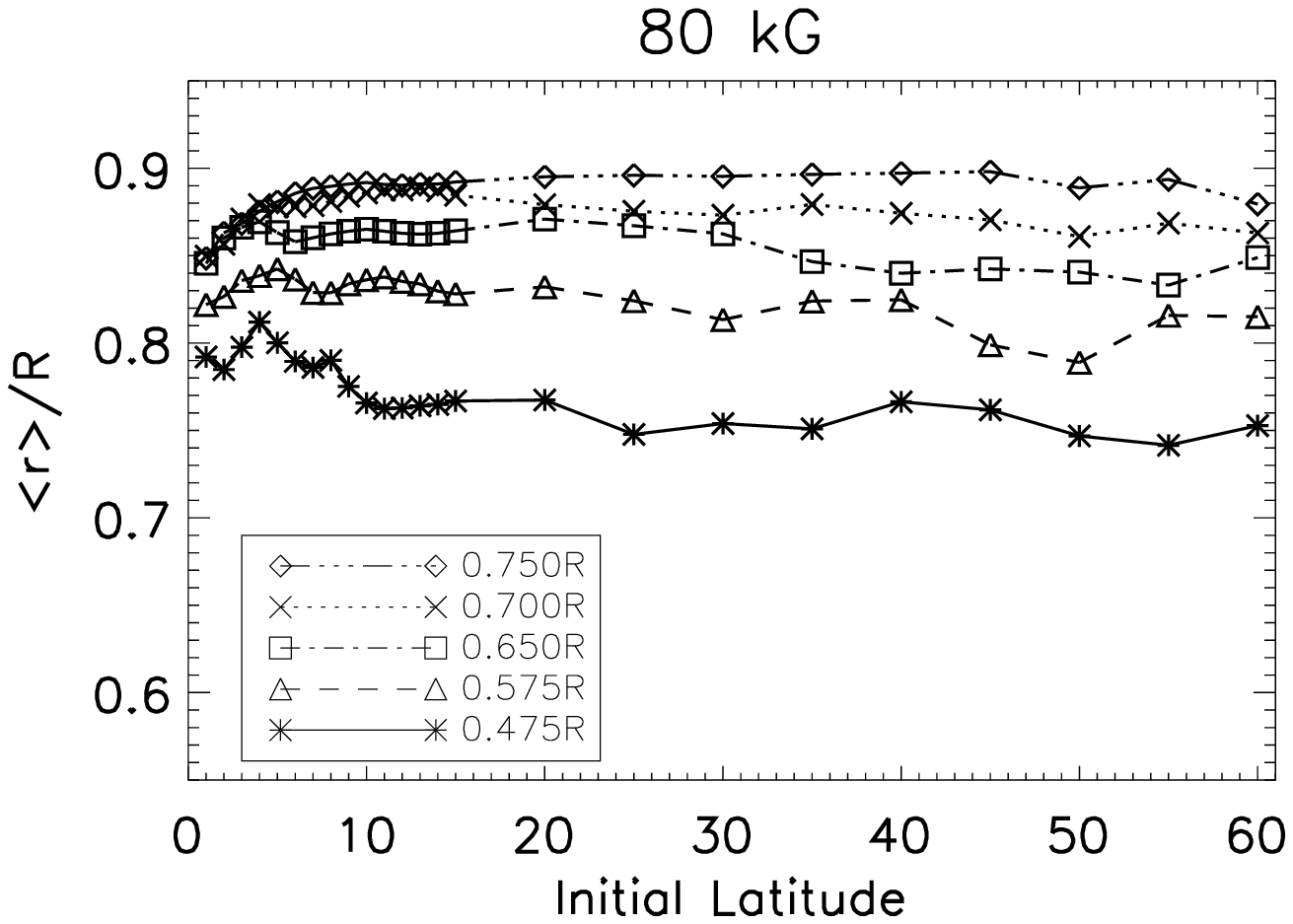}

 \caption{Average magnetic field weighted radial depth $\langle r \rangle$ as of function of $|\theta_{0}|$ for (a) 30 kG and (b) 80 kG flux tubes originating at various depths. This quantity is calculated once the apex of the tube has reached the simulation upper boundary. At this instant, the majority of the flux tube still resides in deeper layers. For any given depth, tubes of weaker magnetic field strengths are also largely confined to deeper layers. Both plots have a different y-axis range.}
   \label{fig:rbavg}
\end{center}
\end{figure}

Figure \ref{fig:rbavg} shows the average magnetic field weighted radial depth $\langle r \rangle$ for 30 kG and 80 kG flux tubes calculated once the apex has reached 0.95R. Near depths of $\sim$0.5R, the critical field strength at which the magnetic buoyancy is roughly equivalent to the drag force from radial convective downflows is $B_{c}$$\sim$30 kG. Even at field strength values close to $B_{c}$, $\langle r \rangle$ increases with time owing to the unbalanced vertical and poleward force components acting on the entire flux tube. As such, even 30 kG tubes initiated at 0.475R reach values of $\langle r \rangle$$\sim$0.6R before the tube apex reaches the surface. As one might expect, weaker $B_{0}$ flux tubes are effectively pinned to deeper layers by downflows than stronger $B_{0}$ tubes originating at the same depth. For any given initial depth, even though rise times may vary greatly over the latitudinal range (see Fig. \ref{fig:rise_times}), the quantity $\langle r \rangle$ does not fluctuate by much more than $5\%$ of the stellar radius.

The degree to which $\langle r \rangle$ deviates in time from the radial position $r_{q}$ of the corresponding flux tube rising in a quiescent interior is an expression of the effectiveness of  `magnetic pumping'. Here we do not refer to magnetic pumping in the traditional sense (e.g. \cite[Tobias et al. 2001]{tobias2001}), but rather to the general ability of convective downflows to suppress the rise of buoyant flux tubes. In some sense, many flux tubes subjected to time-varying flows are `pumped' downwards relative to those that travel through a convection zone void of fluid motions. To capture this process in a single value, we calculate a \emph{relative pumping depth} for each individual flux tube $\Delta r = \langle r(t_{min}) \rangle -r_{q}(t_{min})$. The value $t_{min}$ is the elapsed rise time corresponding to whichever flux tube reaches the upper boundary at 0.95R first, either the flux tube evolving without convective motions or the tube advected by fluid flows (see also \cite[WB16]{wb2016}).     

\begin{figure}
\begin{center}

 \vspace{0.0\textwidth}   % Shift close to the panel top 
     \centerline{\normalsize     % Includes the labels (here needs the color 
                                %   package, see beginning of this file)
      \hspace{0.035 \textwidth}  {(a)}
      \hspace{0.435\textwidth}  {(b)}
         \hfill}
     \vspace{-0.03\textwidth}    % Shift back to the panel bottom

 \includegraphics[scale=.492,clip=true,trim=.8cm 0.5cm .7cm 0cm]{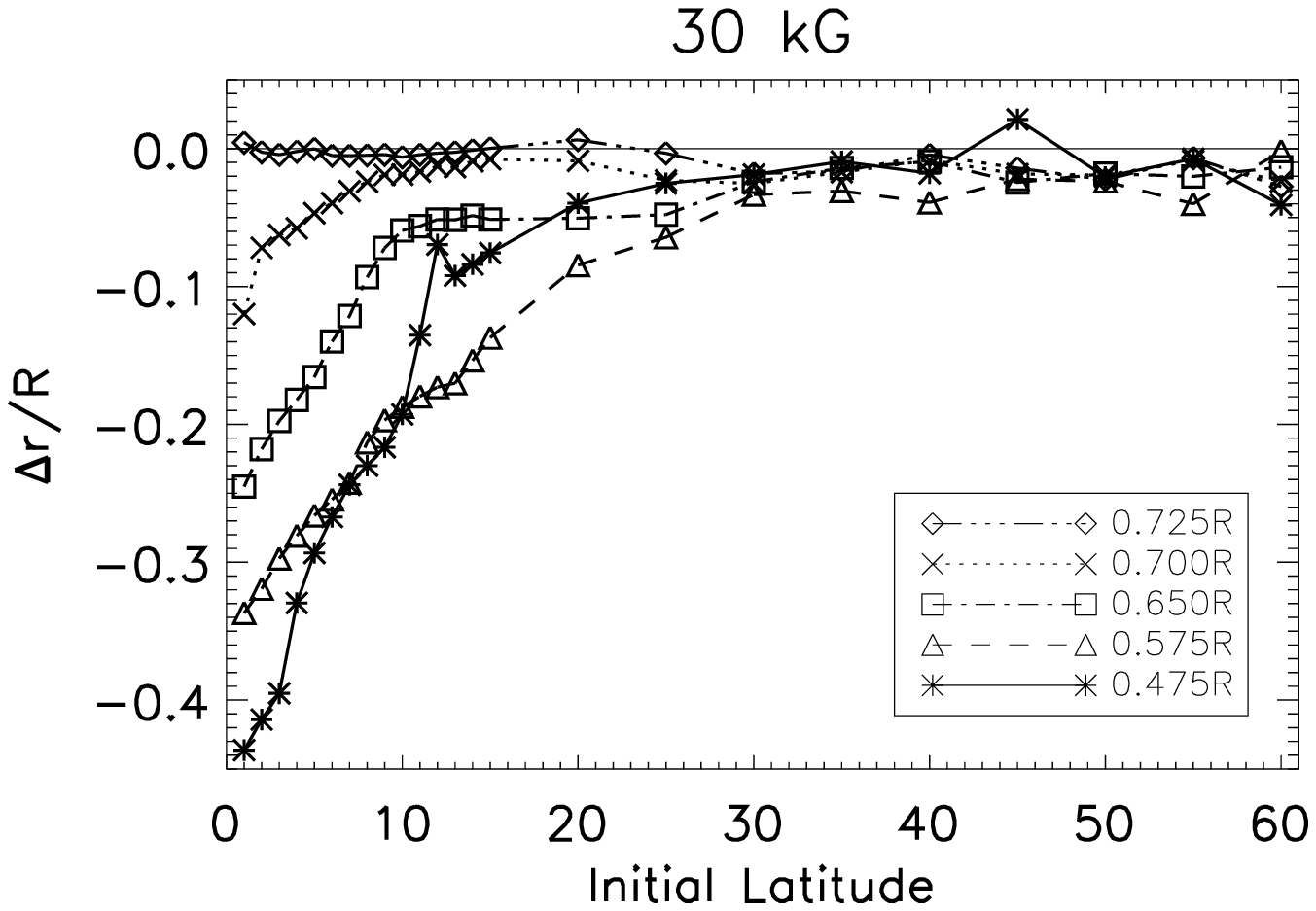} 
 \includegraphics[scale=.492,clip=true,trim=1.05cm 0.5cm .7cm 0cm]{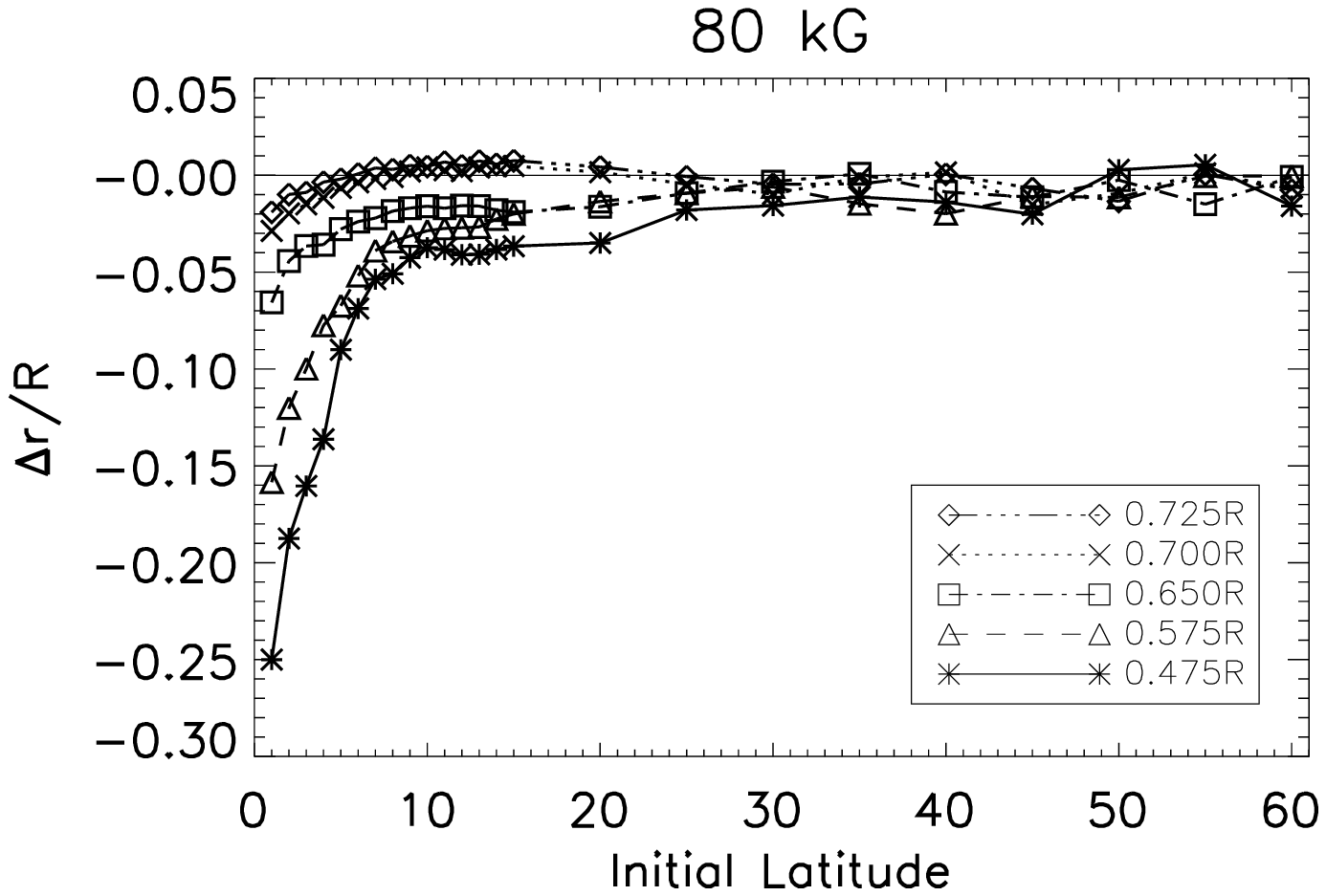}
 
 \caption{Relative pumping depth $\Delta r$ as of function of $|\theta_{0}|$ for (a) 30 kG and (b) 80 kG flux tubes originating at various depths, some of which are different from Fig. \ref{fig:rbavg}. Suppression of the global motion of the flux tube by convective motions is more pronounced at lower latitudes and in the deeper interior. Both plots have a different y-axis range.}
   \label{fig:deltar}
\end{center}
\end{figure}

Figure \ref{fig:deltar} shows the relative pumping depth $\Delta r$ for 30 kG and 80 kG flux tubes originating at various depths. A negative $\Delta r$ indicates that the overall rise of the tube represented by $\langle r \rangle$ is suppressed compared to the flux tube rising in the quiescent interior at the same instant. Even though the majority of flux tubes launched from the same depth attain a similar $\langle r \rangle$, it is clear from Figure \ref{fig:deltar} that the mean motion of tubes initiated at lower latitudes and in deeper layers is strongly suppressed by fluid motions. This suppression of flux emergence reduces as the magnetic field strength increases. At higher latitudes, the mean motion of the flux tube evolves similarly to those that rise through a quiescent convection zone (i.e. smaller $\Delta r$). This is in part a consequence of the larger poleward acceleration due to a greater magnetic tension force arising from a smaller flux tube radius of curvature. Convective motions are not sufficient enough to reduce this motion as effectively. Finally, for the parameter space we explore here, flux tubes of all $B_{0}$ originating in layers $>$0.7R show very little suppression of the mean motion by convection.

\section{Conclusions and Perspectives}
We have reported some results from thin flux tube simulations embedded in a rotating spherical domain of mean and time-varying flows depictive of a 0.3M$_{\odot}$ fully convective star. These simulations model how coherent bundles of magnetic fields, assumed to be starspot progenitors, may behave as they travel through the stellar interior. In this work, we focus on studying how differential rotation and time-varying flows can either promote or suppress the rise of magnetism.    

The archetype of loop-shaped, buoyantly rising magnetic structures is not achievable in the quiescent interior of a fully convective star, assuming the toroidal flux tube is built in thermal equilibrium with its surroundings.  However, when subjected to a time-varying convective velocity, the flux tube develops small undulations that grow into peaks and troughs, promoting rising loops (Sec. \ref{sec:conv_effects}). To a zeroth order approximation, these flux tubes rise parallel to the rotation axis. Yet, flux tubes of a few times $B_{c}$ (where $B_{c}$$\sim$20-30 kG), initiated at lower latitudes and in shallower depths of $\gtrsim$0.725R begin to exhibit near-equatorial emergence and rise times significantly shorter than expected (Sec. \ref{sec:rise_and_lats}). Such behavior arises because the strong prograde differential rotation in this region supplies angular momentum to the legs of the rising loop. When initiated in the deep interior ($\sim$0.475-0.65R), these same flux tubes exhibit significantly longer rise times than their counterparts rising in a quiescent convection zone (Sec. \ref{sec:rise_and_lats}). The apices of these tubes are continually pummeled by downflows, which also act to retard the mean motion of the flux tube. A broad trend emerges: the overall suppression of flux emergence by convective motions is more effective for weaker magnetic field strengths initiated at lower latitudes in deeper layers (Sec. \ref{sec:mag_pump}). 

These TFT simulations complement 3D dynamo simulations of fully convective stars. They inform us about the processes at work in stellar interiors that may lead to the observed pattern of magnetism on stellar surfaces. They also guide our understanding of the relative timescales over which flux emergence may occur and how the majority of this magnetic field may be effectively retained within a star. We plan to further explore these topics by performing similar TFT simulations in a variety of M dwarfs, some of which may have rapid rotation or small radiative cores.     

This work was supported by the ERC under grant agreement No. 337705 (CHASM) and by a Consolidated Grant from the UK STFC (ST/J001627/1). Some calculations were performed on DiRAC Complexity - jointly funded by STFC and the Large Facilities Capital Fund of BIS, and the University of Exeter supercomputer - a DiRAC Facility jointly funded by STFC, the Large Facilities Capital Fund of BIS and the University of Exeter. We thank Isabelle Baraffe for providing the 1D stellar structure model used in this work. Figure \ref{fig:flow_and_tube}b was generated by VAPOR (\cite[Clyne et al. 2007]{clyne2007}). We acknowledge PRACE for awarding us access to computational resources Mare Nostrum based in Spain at the Barcelona Supercomputing Center, and Fermi and Marconi based in Italy at Cineca. S.B, J.C., S.P, and E.T. are MPhys students in the Dept. of Physics and Astronomy at the University of Exeter.

%\begin{discussion}
%\end{discussion}

\end{document}